\def\year{2022}\relax
\documentclass[letterpaper]{article} 
\usepackage{aaai22}  
\usepackage{times}  
\usepackage{helvet}  
\usepackage{courier}  
\usepackage[hyphens]{url}  
\usepackage{graphicx} 
\usepackage{natbib}  
\usepackage{caption} 
\DeclareCaptionStyle{ruled}{labelfont=normalfont,labelsep=colon,strut=off} 
\usepackage[ruled, vlined]{algorithm2e}
\SetKwInOut{Parameter}{parameter}
\usepackage[]{hyperref}

\usepackage{multirow}
\usepackage{enumerate}
\usepackage[switch]{lineno}
\usepackage{paralist}
\usepackage{mathrsfs}
\usepackage{amssymb}
\usepackage{amsmath}
\usepackage{dsfont}
\usepackage{cleveref}
\usepackage{makecell}
\usepackage{threeparttable}
\usepackage{diagbox}
\usepackage{pifont}
\usepackage{relsize}

\usepackage{amsfonts}
\usepackage{booktabs}

\usepackage{pifont}
\usepackage{paralist}
\usepackage{enumerate}
\usepackage{relsize}

\usepackage{xcolor}
\usepackage{graphicx}
\usepackage{amssymb}
\usepackage{amsmath}

\usepackage{subfigure}

\usepackage{framed,multirow}

\def\spac{\texttt{SPAC}}

\makeatletter
\newcommand{\printfnsymbol}[1]{%
  \textsuperscript{\@fnsymbol{#1}}%
}
\makeatother

\title{Stochastic Planner-Actor-Critic for Unsupervised Deformable Image Registration}
\author {
    Ziwei Luo\textsuperscript{\rm 1}\thanks{Equal contribution.},
    Jing Hu\textsuperscript{\rm 1}\printfnsymbol{1},
    Xin Wang\textsuperscript{\rm 2}\thanks{Corresponding authors.},
    Shu Hu\textsuperscript{\rm 3},\\
    Bin Kong\textsuperscript{\rm 2},
    Youbing Yin\textsuperscript{\rm 2},
    Qi Song\textsuperscript{\rm 2},
    Xi Wu\textsuperscript{\rm 1}\printfnsymbol{2},
    Siwei Lyu\textsuperscript{\rm 3}
}
\affiliations {
    \textsuperscript{\rm 1} Chengdu University of Information Technology, China\\
    \textsuperscript{\rm 2} Keya Medical, Seattle, USA\\
    \textsuperscript{\rm 3} University at Buffalo, SUNY, USA\\
}

\begin{document}

\maketitle

\begin{abstract}
Large deformations of organs, caused by diverse shapes and nonlinear shape changes, pose a significant challenge for medical image registration. Traditional registration methods need to iteratively optimize an objective function via a specific deformation model along with meticulous parameter tuning, but which have limited capabilities in registering images with large deformations. While deep learning-based methods can learn the complex mapping from input images to their respective deformation field, it is regression-based and is prone to be stuck at local minima, particularly when large deformations are involved. To this end, we present Stochastic Planner-Actor-Critic (\spac), a novel reinforcement learning-based framework that performs step-wise registration. The key notion is warping a moving image successively by each time step to finally align to a fixed image. Considering that it is challenging to handle high dimensional continuous action and state spaces in the conventional reinforcement learning (RL) framework, we introduce a new concept `Plan' to the standard Actor-Critic model, which is of low dimension and can facilitate the actor to generate a tractable high dimensional action. The entire framework is based on unsupervised training and operates in an end-to-end manner. We evaluate our method on several 2D and 3D medical image datasets, some of which contain large deformations. Our empirical results highlight that our work achieves consistent, significant gains and outperforms state-of-the-art methods. Code is available at {\url{https://github.com/Algolzw/SPAC-Deformable-Registration}}
\end{abstract}

\section{Introduction}

Deformable image registration (DIR) is an important task in medical imaging 
and has been actively studied for decades. DIR consists of establishing a spatial anatomical non-linear dense correspondence between a pair of fixed and moving images. 
The central task of DIR is the estimation of the ill-posed free-form transformation field that consists of a combination of global and local displacements \cite{eppenhof2019progressively}. 
Moreover, an accurate DIR on large deformation is needed due to soft organs (e.g., the brain, liver, and stomach) may undergo large deformations caused by patient re-positioning, surgical manipulation, or other physiological differences \cite{holden2007review}. 
Large deformation diffeomorphic metric mapping (LDDMM) \cite{beg2005computing}, derived from the group structure of the manifold of diffeomorphisms, is one of the most popular methods to tackle large deformations in DIR. 
However, achieving an optimal solution of the diffeomorphic image registration is computationally intensive and time-consuming, attempts at speeding up diffeomorphic image registration have thus been proposed to improve numerical approximation schemes \cite{wang2020deepflash}. 

Most existing DL-based solutions \cite{balakrishnan2019voxelmorph,dalca2019unsupervised,mok2020fast} are enforced to make a straightforward prediction, which is incapable to handle complicated deformations \cite{zhao2019recursive}. 
The step-wise image registration methods, such as R2N2 \cite{NEURIPS2019_dd03de08} and RCN \cite{zhao2019recursive}, have shown the potential in DIR, in which the final deformation field (probably with large displacements) can be considered as a composition of the progressively predicted deformation field. But both of them are complex and computationally costly, and cannot deal with long step-wise registration.
Inspired by the way that a human expert aligns two images by applying a sequence of local or global deformations,
some RL-based image registration methods have been introduced in the past \cite{liao2017artificial,ma2017multimodal,miao2019agent,sun2018robust,hu2021end,luo2020spatiotemporal}. However, most of them merely focus on global rigid transformation since it only includes rotation and translation and can be easily represented by a low-dimensional discrete parametric model.
Compared to rigid registration, DIR has huge, continuous state and action spaces, especially in 3D, which makes RL training extremely difficult. 
Krebs et al.~\cite{krebs2017robust} proposed an RL-based approach for DIR, but their method uses supervised learning with ground truth deformation field, which is infeasible for most DIR tasks. And they incorporate the traditional statistical deformation model to reduce and discretize the action space, which leads to inferior performance in complex deformation registration. 

In this paper, we propose a new RL architecture for unsupervised DIR problems, known as the {\em Stochastic Planner-Actor-Critic} (\spac), to handle high dimensional continuous state and action spaces.
As shown in Figure \ref{overview}, the \spac~framework is formed with three core deep neural networks: the planner, the actor, and the critic.
We introduce new a concept `plan' which breaks the decision-making process into two steps, state $\rightarrow$ plan and plan $\rightarrow$ action. We call this process as meta policy, where the plan  
is a subspace of appropriate actions based on the current state, but it is not applied to the state directly, it is used to guide the actor to generate a tractable high-dimensional action that applies to the environment. The plan could be considered as an intermediate transition between state and action. As the input of the actor, the plan has a much lower dimension comparing with the state, which is easier for the actor to learn to predict actions. Meanwhile, the plan can be evaluated by the critic efficiently, since the Q function is easier to learn in the low-dimensional latent space.
Furthermore, we employ an unsupervised registration learning strategy to learn the similarity of appearance between fixed and moving image pairs.
The main contributions of our work can be summarized as follows:
\begin{itemize}
\item We describe a new RL framework, stochastic planner-actor-critic (\spac), to handle large deformations by decomposing the monolithic learning process in DIR into small steps with high-dimensional continuous actions.
\item To tackle the high-dimensional continuous action learning problem, we propose a stochastic meta policy that breaks the decision-making processing into two steps: state $\rightarrow$ low-dimensional plan and plan $\rightarrow$ deformation field action. The plan guides the actor to predict a tractable action, and the critic evaluates the plan, which makes the whole learning process feasible and computationally efficient.
\item We design an registration environment which incorporates a K-means clustering module \cite{dice1945measures} to obtain coarse segmentation maps to compute the Dice reward in an unsupervised manner, which obviates the need to collect real data with abundant and reliable ground-truth annotations. Besides, our method can be applied to entire 3D volumes.
\item Experimental results on a variety of 2D/3D datasets show that the \spac~achieves state-of-the-art performance and consistently improves the results along with iterations. 
\end{itemize}


\section{Background}
\label{sec:bg}
\subsection{Deformable Image Registration}
Deformable image registration (DIR) aims to learn a transformation between a fixed image and a moving image \cite{yang2016fast,balakrishnan2018unsupervised,de2019deep}. 
The DIR can be defined as an optimization problem. Given a pair of images ($I_F, I_M$), both of which on the image domain ${\cal X} \rightarrow \mathbb{R}^d$, where $d$ is the dimension. $I_F$ is the fixed image and $I_M$ is the moving image. Denote $\Omega_{w}$ as a registration model parameterized by $w$. The output of it is a deformation filed, which can be warped on the moving image to align to the fixed image, denoted as $I_M \circ \Omega_{w}(I_F, I_M)$. We formulated the pairwise registration as a minimization problem based on energy function:

\begin{equation}
\min_w E(w) := G(I_F, I_M\circ \Omega_w(I_F, I_M))+\lambda R(\Omega_w(I_F, I_M))
\label{eqDIR}
\end{equation}
where $G$ represents a metric quantifying the similarity between the fixed image and the warped image, 
$R$ represents a regularization
constraining the deformation field, $\lambda$ is a hyperparameter to balance these two terms. In the DL-based DIR task, the DL model tries to learn $\Omega_w(I_F, I_M)$ from a training dataset, which contains a large number of image pairs ($I_F, I_M$). 
The potential choice of $G$ could be any similarity metric, such as the sum of squared differences (SSD), the normalized mutual information (NMI), or the negative normalized cross-correlation (NCC) \cite{haskins2020deep, balakrishnan2018unsupervised}.

\subsection{Reinforcement Learning} 
RL is described by an infinite-horizon Markov decision process (MDP), defined by the tuple $({\cal S},{\cal A}, {\cal U},{r}, {\gamma})$. ${\cal S}$ is a set of states, ${\cal A}$ is action, and ${\cal U}: {\cal S}  \times {\cal S} \times {\cal A} \rightarrow [0, \infty)$ represents the state transition probability density given state $s \in {\cal S}$ and action ${\bf a} \in {\cal A}$. ${r}: {\cal S} \times {\cal A} \rightarrow \mathbb{R}$ is the reward emitted from each transition, and $\gamma \in [0,1]$ is the reward discount factor.
Standard RL learns to maximize the expected sum of rewards from the episodic environments under the trajectory distribution ${\rho}_{\pi}$.
It can be modified to incorporate an entropy term with the policy. Therefore, the resulting objective is defined as $\sum \nolimits_{t = 1}^T \mathbb{E}_{({\bf s}_t, {\bf a}_t) \sim {\rho}_{\pi}} \left[r_t ({\bf s}_t, {\bf a}_t) + \alpha {\cal H}(\pi_\phi (\cdot | {\bf s}_t))\right]$, where $\alpha$ is a temperature parameter controlling the balance of the entropy ${\cal H}$ and the reward $r_t$.   

Soft Actor-critic (SAC) \cite{haarnoja2018soft} is a promising framework for learning continuous actions, which is an off-policy actor-critic method that uses the above entropy-based framework to derive the soft policy iteration. Stochastic latent actor-critic (SLAC) improves the SAC by learning the representation spaces with a latent variable model which is more stable and efficient for complex continuous control tasks.
It can improve both the exploration and robustness of the learned model. 
However, SLAC is far from enough for handling DIR, which has huge continuous action spaces such as voxel-wise estimation of a deformation field. 

\section{Stochastic Planner-Actor-Critic}



\begin{figure*}[t]
    \centering
    \includegraphics[scale=0.62]{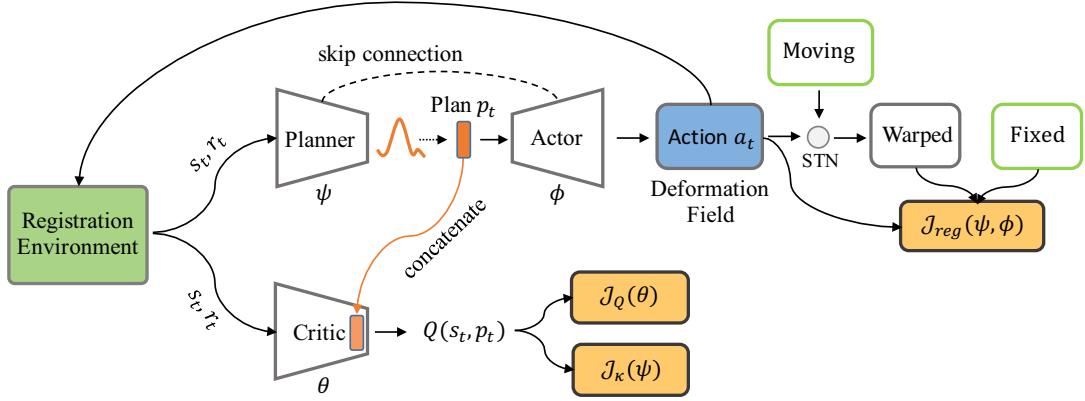}
    \caption{The network architecture of the proposed \spac~ for DIR problem. At time step $t$, the registration environment receives action ${\bf a}_t$, and outputs state and reward ($s_t, r_t$). The plan ${\bf p}_t$ is sampled from the planner and evaluated by the critic. The action ${\bf a}_t$ is actually an immediate deformation field based on state $s_t$. The spatial transformer network (STN) is used as the warping function.}
    \label{overview}
\end{figure*}

Concretely, in our framework, the \spac~is formed with three core deep neural networks: the planner, the actor, and the critic with parameters $\psi$, $\phi$, and $\theta$, respectively (see Figure \ref{overview}). The planner aims to generate a high-level plan in the low-dimensional latent space to guide the actor. In some sense, the plan can be considered as action clusters or action templates, which are high-level crude actions. Different from classic actor-critic models, the input of the actor is a stochastic plan instead of the state. That is, the generated plan is forwarded to the actor to further create the high dimensional action for DIR, and meanwhile, this plan is evaluated by the critic. We also add skip-connections from each down-sampling layer of the planner to the corresponding up-sampling layer of the actor. The information passed by skip-connections contains the details of the state ${\bf s}_t$ that are needed to reconstruct a natural-looking image.
Using the proposed stochastic planner-actor-critic structure and supervision of similarity of appearance between fixed and moving image, \spac~could extend readily to complex DIR tasks.

\subsection{Problem Formulation}

In forming \spac, we apply the same notations as we defined for the conventional reinforcement learning in background section and introduce an additional component ${\cal P}$ for the continuous plan space to the infinite-horizon Markov decision process (MDP). Therefore, the MDP for \spac~can be defined by the tuple $({\cal S},{\cal P},{\cal A}, {\cal U},{r}, {\gamma})$. ${\cal S}$ is a set of states, ${\cal P}$ is continuous plan, ${\cal A}$ is continuous action, and ${\cal U}: {\cal S} \times {\cal P}  \times {\cal S} \times {\cal A}  \rightarrow [0, \infty)$ represents the state transition probability density of the next state $s_{t+1}$ given state ${\bf s}_t \in {\cal S}$, plan $p_t \in {\cal P}$ and action ${\bf a}_t \in {\cal A}$. 

\subsubsection{Step-wise Deformable Registration.} 

Leveraging the sequential characteristic of reinforcement learning, we decompose the registration into $T$ steps instead of predicting the deformation field in one-shot. At time step $t$, action ${\bf a}_t$ is the current deformation field generated by the Planner $\kappa_{\psi}$ and Actor $\pi_{\phi}$ based on fixed image $I_F$ and intermediate moving image $I_{M_t}$. Let $\Omega^t_{\psi, \phi}$ represents the accumulated deformation field composed by ${\bf a}_t$ and the previous deformation field $\Omega^{t-1}_{\psi, \phi}$. We can compute $\Omega^t_{\psi, \phi}$ with a recursive composition function:

\begin{equation}
\begin{aligned}
\Omega^t_{\psi, \phi} =
\begin{cases}
     0 & \mbox{if $t=0$,}\\
     {\cal C}({\bf a_t}, \Omega^{t-1}_{\psi, \phi}) & \mbox {otherwise,}
\end{cases} 
\end{aligned}
\label{eq:compose_field}
\end{equation}
where

\begin{equation}
     \ {\cal C}({\bf a_t}, \Omega^{t-1}_{\psi, \phi}) = \Omega^{t-1}_{\psi, \phi} + ({\bf a}_t \circ \Omega^{t-1}_{\psi, \phi}).
\end{equation}
To eliminate the warping bias in the multi-step recursive registration process \cite{zhao2019recursive}, we warp the initial moving image $I_M$ using accumulated the deformation field $\Omega^t_{\psi, \phi}$.
Then the warped image $I_{M_{t+1}}$ is used as the next moving image in time step $t+1$.  
Therefore, the registration result can be progressively improved by predicting deformation from coarse to the local refined. 
Using the notion of step-wise, the DIR optimization problem (Eq.(\ref{eqDIR})) in our \spac~framework can be rewritten as

\begin{equation}
    \begin{aligned}
    \min_{\psi,\phi} E({\psi,\phi}) := \frac{1}{T} \sum_{t=1}^T G(I_F, I_{M_t}\circ \Omega^t_{\psi,\phi})+\lambda R(\Omega^t_{\psi,\phi}),
    \end{aligned}
    \label{eq:DIR}
\end{equation}
where we use a tuple ($\psi, \phi$) instead of the parameter $w$ in Eq.(\ref{eqDIR}) since the deformation field will be learned from the \spac~framework. In the following, we provide the details of the environment and policy in our method.

\subsection{DIR Environment} 
\begin{figure}[t]
    \centering
    \includegraphics[scale=0.7]{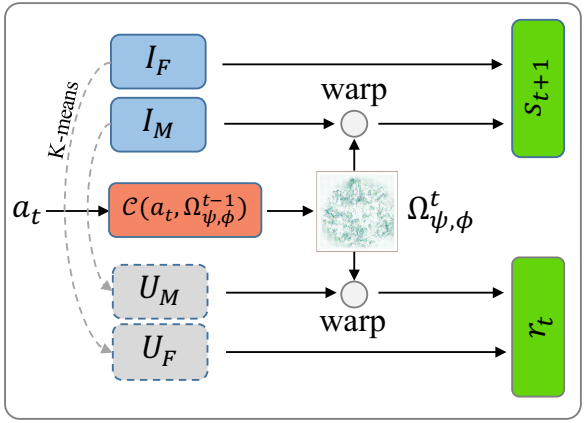}
    \caption{The architecture of the step-wise deformable registration environment. When the environment receives an action ${\bf a}_t$, it outputs the next state $s_{t+1}$ and reward $r_t$. Specifically, the environment is composed by a pair of image ($I_F$, $I_M$), and generates the corresponding segmentation maps ($U_F$, $U_M$) with K-means clustering. ${\cal C}(a_t, \Omega^{t-1}_{\psi, \phi})$ is the compose function which composes action $a_t$ to the accumulated deformation field $\Omega^t_{\psi, \phi}$. The next state $s_{t+1}$ is obtained by concatenating $I_F$ and the warped moving image $I_M \circ  \Omega^t_{\psi, \phi} $. The reward $r_t$ is obtained by Eq. (\ref{eq:reward}) which represents the improvement of Dice score.}
    \label{fig:env}
\end{figure}

The overview of the step-wise deformable registration environment is shown in Figure \ref{fig:env}. In the beginning, the environment only contains an image pair ($I_F$, $I_M$), then we performs K-means \cite{macqueen1967some} with three clustering labels to obtain the corresponding segmentation maps ($U_F$, $U_M$) in an unsupervised manner. The generated segmentation map assigns each voxel to a virtual anatomical structure, which facilitates computing rewards. At time step $t$, the state ${\bf s}_t$ is the pair of fixed image $I_F$ and moving image $I_{M_t}$, ${\bf s}_t=(I_F, I_{M_t})$. The next state $s_{t+1}$ is obtained by warping $I_{M}$ with composed deformable field $\Omega^t_{\psi,\phi}$: $s_{t+1}=(I_F, I_M \circ \Omega^t_{\psi,\phi})$. 
We incorporate the widely used spatial transformer network (STN) \cite{jaderberg2015spatial} as the warping operator.
The reward ${r_t}$ is defined based on the {\rm Dice} \cite{dice1945measures} score: 

\begin{equation}
r_t = {\rm Dice}(U_F,  U_M \circ \Omega^t_{\psi,\phi}) - {\rm Dice}(U_F,  U_M \circ \Omega^{t-1}_{\psi,\phi}),
\label{eq:reward}
\end{equation}
where ${\rm Dice}(U_1, U_2) = 2 \cdot \frac{|U_1 \cap U_2|}{|U_1| + |U_2|}$. This reward function explicitly assesses the improvement of the predicted deformation field $\Omega^t_{\psi,\phi}$. 

\subsection{Stochastic Meta Policy}
\label{sec:metaplicy}
In our formulation, we have a meta policy ($\kappa,\pi$), where the stochastic plan is modeled as a subspace of the deformation field that gives low-dimensional vector ${\bf p}_t$ given state ${\bf s}_t$. While the actor's action is actually a deterministic deformation field ${\bf a}_t$ determined by the plan ${\bf p}_t$. 
Consider a parameterized planner $ \kappa_{\psi}$ and actor $\pi_{\phi}$, the stochastic plan is sampled as a representation: ${\bf p}_t \sim \kappa_{\psi}( {\bf p}_t|{\bf s}_t)$, and the action is generated by decoding the plan vector ${\bf p}_t$ to a high-dimensional deformation field: ${\bf a}_t = \pi_{\phi}( {\bf a}_t|{\bf p}_t)$  In practice, we reparameterize the the planner and the stochastic plan jointly using a neural network approximation ${\bf p}_t = f_\psi({\bf \epsilon}_t,{\bf s}_t)$, known as reparameterization trick~\cite{kingma2013auto}, where ${\bf \epsilon}_t$ is an input noise vector sampled from a fixed Gaussian distribution. Moreover, we maximize the entropy of plan to improve exploration and robustness. The augmented objective function is formulated as:

\begin{equation*}
    \begin{aligned}
  \max_{\psi,\phi} \sum \limits_{t = 1}^T \mathbb{E}_{({\bf s}_t, {\bf p}_t,{\bf a}_t) \sim {\rho}_{(\kappa,\pi)}} \left[r_t ({\bf s}_t,{\bf p}_t, {\bf a}_t) + \alpha {\cal H}(\kappa_\psi (\cdot | {\bf s}_t))\right]. 
    \end{aligned}
\end{equation*}
where $\alpha$ is the temperature and ${\rho}_{(\kappa,\pi)}$ is a trajectory distribution under $\kappa_{\psi}( {\bf p}_t|{\bf s}_t)$ and $\pi_{\phi}( {\bf a}_t|{\bf p}_t)$. 

\subsection{Learning Planner and Critic }

Different from conventional RL algorithms, the critic $Q_{\theta}$ evaluates plan ${\bf P}_t$ instead of action ${\bf a}_t$.
since learning a low-dimensional plan in the DIR problem is easier and more effective. 
Specifically, the low-dimensional plan is concatenated to the downsampled vector of the critic and outputs soft Q function $Q_{\theta}({\bf s}_t, {\bf p}_t)$ which is an estimation of the current state plan value, as shown in Figure \ref{overview}.

When the critic is used to evaluate the planner, the rewards and the soft Q values are used to iteratively guide the stochastic policy improvement. 
In the evaluation step, following SAC \cite{haarnoja2018soft}, \spac~learns a policy $\kappa_\psi$ (planner) and fits the parametric Q-function $Q_{\theta}({\bf s}_t,{\bf p}_t)$ (critic) using transitions sampled from the replay pool $\mathcal{D}$ by minimizing the soft Bellman residual:

\begin{equation*}
    \begin{aligned}
    &J_Q(\theta) = \\
    &\mathbb{E}_{({\bf s}_t,{\bf p}_t)\sim  \mathcal{D}} \left[\frac{1}{2} \Big(Q_{\theta}({\bf s}_t, {\bf p}_t) - \big(r_t + \gamma \mathbb{E}_{{\bf s}_{t+1}}\left[V_{\bar{\theta}}({\bf s}_{t+1})\right]\big)\Big)^2\right],
    \end{aligned}
\end{equation*}
where $V_{\bar{\theta}}({\bf s}_{t}) = \mathbb{E}_{{\bf p}_{t}\sim \kappa_\psi}[Q_{\bar{\theta}}({\bf s}_{t}, {\bf p}_{t}) - \alpha \log \kappa_{\psi} ({\bf p}_{t}|{\bf s}_{t})]$. We use a target network $Q_{\bar{\theta}}$ to stabilize training, whose parameters $\bar{\theta}$ are obtained by an exponentially moving average of parameters of the critic network \cite{lillicrap2015continuous}: $\bar{\theta} \rightarrow \tau \theta + (1-\tau)\bar{\theta}$. The hyper-parameter $\tau\in [0,1]$. To optimize the $J_Q(\theta)$,  we can do the stochastic gradient descent with respect to the parameters $\theta$ as follows,

\begin{equation}
    \begin{aligned}
    \theta = \theta &- \eta_Q \triangledown_{\theta} Q_{\theta}({\bf s}_t, {\bf p}_t)\Big(Q_{\theta}({\bf s}_t, {\bf p}_t) - r_t \\ 
    &- \gamma \left[Q_{\bar{\theta}}({\bf s}_{t+1}, {\bf p}_{t+1}) - \alpha \log \kappa_{\psi} ({\bf p}_{t+1}|{\bf s}_{t+1})\right]\Big).
    \end{aligned}
\label{eq:update_theta}
\end{equation}
Since the critic works on the planner, the optimization procedure will also influence the planner decisions.
Following \cite{haarnoja2018soft}, we can use the following objective to minimize the KL divergence between the policy and a Boltzmann distribution induced by the Q-function,

\begin{equation*}
    \begin{aligned}
    J_{\kappa} (\psi) =& \mathbb{E}_{ {\bf s}_t\sim  \mathcal{D}}\big[ \mathbb{E}_{{\bf p}_t\sim  \kappa_{\psi}} \left[\alpha \log (\kappa_{\psi}({\bf p}_t| {\bf s}_t))-Q_\theta({\bf s}_t,{\bf p}_t)\right]\big]\\
    =&\mathbb{E}_{ {\bf s}_t\sim  \mathcal{D}, {\bf \epsilon}_t\sim  \mathcal{N} (\mu,\sigma) } \big[\alpha \log (\kappa_{\psi}( f_\psi({\bf \epsilon}_t,{\bf s}_t)|{\bf s}_t))\\
    &\phantom{=\;;;;;;;;;;;;;;;;;;;;;;;;}
    - Q_\theta({\bf s}_t,f_\psi({\bf \epsilon}_t,{\bf s}_t))\big].
    \end{aligned}
\end{equation*}
The last equation holds because ${\bf p}_t$ can be evaluated by $f_\psi({\bf \epsilon}_t,{\bf s}_t)$ as we discussed before. It should be mentioned that the hyperparameter $\alpha$ can be automatically adjusted by using one proposed method from \cite{haarnoja2018soft}. Then we can apply the stochastic gradient method to optimize parameters as follows,

\begin{equation}
    \begin{aligned}
    \psi = &\psi -\eta_\psi\Big(\triangledown_\psi \alpha \log(\kappa_\psi(p_t|s_t)) + \\
    &\big(\triangledown_{p_t}\alpha \log(\kappa_{\psi}(p_t|s_t))-\triangledown_{p_t}Q_{\theta}(s_t,p_t)\big) \triangledown_\psi f_\psi (\epsilon_t,s_t)\Big).
    \end{aligned}
\label{eq:update_psi}
\end{equation}

\subsection{Learning Planner and Actor with Unsupervised Registration}


After getting action ${\bf a}_t$ and following meta policy ($\kappa_{\psi}, \pi_{\phi}$), we can obtain $\Omega^t_{\psi,\phi}$ based on Eq.(\ref{eq:compose_field}). 
We learn the similarity of appearance between the fixed image and the warped image using local normalized cross-correlation (NCC)~\cite{balakrishnan2018unsupervised}: $G(I_F, I_{M_t})=NCC(I_F, I_M \circ \Omega^t_{\psi, \phi})$. A higher value of the NCC indicates a better alignment. 
\begin{figure*}[t]
    \centering
    \includegraphics[scale=0.58]{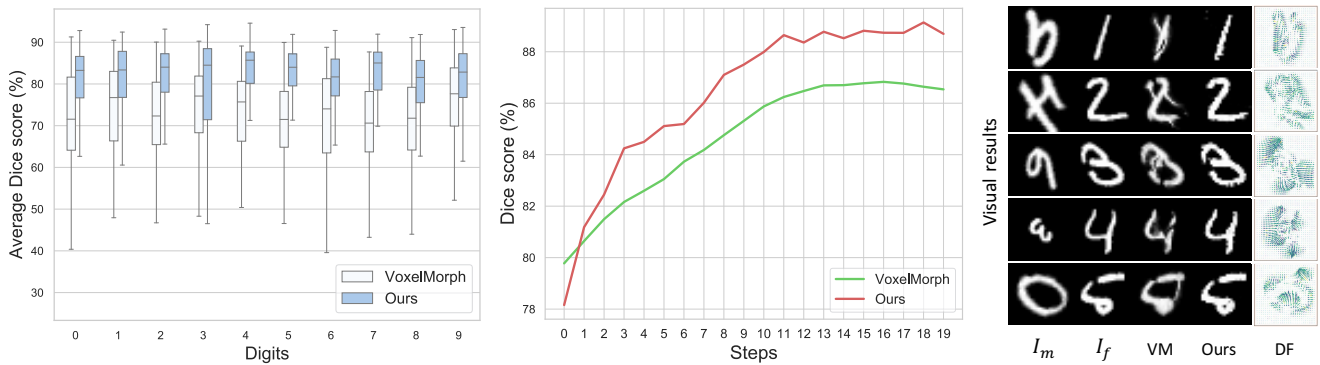}
    \caption{\textbf{Left:} The plot box of Dice scores over 9 fixed digits. \textbf{Center:} Step-wise comparison of our method and VoxelMorph (VM) \cite{balakrishnan2018unsupervised}. \textbf{Right:}  Visual comparison of our method with VM. The scaled and rotated digits are transformed to other fixed digits. The Deformation Filed (DF) column shows the visualized deformable fields of our method.}
    \label{mnist_box}
\end{figure*}
In order to generate realistic warped images, we use a total variation regularizer~\cite{rudin1992nonlinear} to smooth the deformation field on its spatial gradients: $R(\Omega^t_{\psi,\phi}) = \| \nabla \Omega^t_{\psi,\phi}\|^2_2$ . The final registration loss $J_{reg}$ is defined as

\begin{equation*}\small
    \begin{aligned}
    J_{reg}  (\psi, \phi) = & \mathbb{E}_{s_t \sim {\cal D}}[-NCC(I_F, I_M\circ \Omega^t_{\psi, \phi})  
     + \lambda \| \nabla \Omega^t_{\psi,\phi}({\bf s}_t) \|^2_2]. \\
    \end{aligned}
\end{equation*}
We can update $\psi$ and $\phi$  from the planner and the actor by performing the following steps: 
 \begin{equation}
    \begin{aligned}
    \psi = \psi-\eta\triangledown_\psi J_{reg}  (\psi, \phi), \ \ \ \phi = \phi-\eta\triangledown_\phi J_{reg}  (\psi, \phi)
    \end{aligned}
    \label{eq:update_double}
\end{equation}
The pseudo-code of optimizing \spac~is described in Algorithm \ref{Alg0}. All parameters of \spac~are optimized base on the samples from replay pool $\cal D$. 

\begin{algorithm}[t]
    \caption{Stochastic Planner-Actor-Critic}\label{Alg0}
    \SetAlgoLined

    \KwIn{$I_F$, $I_M$, $U_F$, $U_M$, replay pool $\mathcal{D}$}
    
    
    \textbf{Init:} $\psi$, $\phi$, $\theta$, $\bar{\theta}$, $\mathcal{D}$ and environment $\cal E$
    
    \For{each iteration}{
    \For{each environment step}{
    ${\bf p}_{t} \sim \kappa_\psi({\bf p}_{t} | {\bf s}_{t})$,
    ${\bf a}_{t} \sim \pi_\phi({\bf a}_{t} | {\bf p}_{t})$
    
    ${\bf s}_{t+1}, {r_t} \sim \mathcal{U}({\bf s}_{t+1}|{\bf s}_{t},  {\bf p}_{t}, {\bf a}_{t})$
    
    $\mathcal{D} = \mathcal{D} \cup \{({\bf s}_{t}, {\bf p}_{t}, {\bf a}_{t}, r_t, s_{t+1})\}$
    }
    
    \For{each gradient step}{
    Sample from $\mathcal{D}$
    
    Update $\theta$, $\psi$, $\phi$ with Eq.(\ref{eq:update_theta}), Eq.(\ref{eq:update_psi}), Eq. (\ref{eq:update_double}) 
    }
    }
\end{algorithm}

\section{Experiments}

\subsection{Experimental Settings}
\subsubsection{Datasets.} We evaluate our method on three types of images: MNIST digits, 2D brain MRI scans, and 3D liver CT scans. MNIST~\cite{lecun1998gradient} is regarded as a standard sanity check for the proposed registration method. The goal is to transform between two different $28 \times 28$ images of handwritten digits. In testing, we fixed ten digits from 0 to 9 as the atlases, and select 1000 randomly scaled and rotated digits as moving images to be aligned with the atlas.

The 2D brain MRI training dataset consists of 2302 pre-processed 2D scans from ADNI \cite{mueller2005ways}, ABIDE \cite{di2014autism} and ADHD \cite{bellec2017neuro}. 
The evaluation dataset uses 40 pre-processed slices from LONI Probabilistic Brain Atlas (LPBA) \cite{shattuck2008construction}, each of which contains a segmentation ground truth of 56 manually delineated anatomical structures. All images are resampled to $128 \times 128$ pixels. The first slice of LPBA is served as the atlas, and all the remaining images are used as the moving image. 
For 3D registration, we use Liver Tumor Segmentation (LiTS) \cite{bilic2019liver} challenge data for training, which contains 131 CT scans with the segmentation ground truth manually annotated by experts. The SLIVER \cite{heimann2009comparison} dataset has 20 scans with liver segmentation ground truth. We divide them into 10 pairs as the regular testing data. We also evaluated our method on the challenging Liver Segmentation of Pigs (LSPIG) \cite{zhao2019recursive} dataset, which contains 17 paired CT scans from pigs, along with liver segmentation ground truth. All 3D volumes are resampled to $128 \times 128 \times 128$ pixels and pre-affined as standard pre-processing steps.

\begin{table*}[t]
\renewcommand\arraystretch{.9}
\centering
\setlength{\abovecaptionskip}{0.05in}
\setlength{\belowcaptionskip}{-0.1in}
\begin{tabular}{l|ccc|cccc}

\toprule
\multirow{2}{*}{Method} & \multicolumn{3}{c|}{2D Registration}   &  \multicolumn{4}{c}{3D Registration} \\ 
& LPBA   & Time(s) & \#Params   &  SLIVER   & LSPIG & Time(s) & \#Params \\ \midrule
 
SyN~\cite{avants2008symmetric} & 55.47$\pm$3.96    & 4.57 & -  &  89.57$\pm$3.34 & 81.83$\pm$8.30 & 269 & - \\

Elastix~\cite{klein2009elastix} & 53.64$\pm$3.97    & 2.20 & - & 90.23$\pm$2.39    & 81.19$\pm$7.47    & 87.0    & - \\

LDDMM~\cite{beg2005computing} & 52.18$\pm$3.48    & 3.27 & - & 83.94$\pm$3.44    & 82.33$\pm$7.14    & 41.4    & - \\

VM~\cite{balakrishnan2019voxelmorph}  & 55.36$\pm$3.94  & 0.02 & 105K  & 86.37$\pm$4.15  & 81.13$\pm$7.28    & 0.13 & 356K \\

VM-diff~\cite{dalca2019unsupervised}  & 55.88$\pm$3.78  & 0.02 & 118K  & 87.24$\pm$3.26    & 81.38$\pm$7.21    & 0.16  & 396K \\

R2N2~\cite{NEURIPS2019_dd03de08} & 51.84$\pm$3.30   & 0.46 & 3,591K   & - & - & - & - \\

RCN~\cite{zhao2019recursive}    &  -  & - & - & 89.59$\pm$3.18    & 82.87$\pm$5.69    & 2.44   & 21,291K \\

SYMNet~\cite{mok2020fast}  & -  & - & -   & 86.97$\pm$3.82  & 82.78$\pm$7.20 & 0.18    & 1,124K \\
\midrule

\spac~($t$=20, SSIM reward)   & 56.43$\pm$3.76  & 0.16 & 107K  &90.27$\pm$3.85     & 83.69$\pm$6.74    & 1.05 & 458K \\
\midrule

\spac~($t$=1, Dice reward)   & 55.21$\pm$3.55  & 0.02 & 107K  & 84.81$\pm$4.42    & 80.61$\pm$7.94    & 0.07 & 458K \\

\spac~($t$=10, Dice reward)    & 56.12$\pm$3.68  & 0.08 & 107K  & 90.01$\pm$3.79    & \textbf{84.67$\pm$6.05}    & 0.55 & 458K \\

\spac~($t$=20, Dice reward)    & \textbf{56.57$\pm$3.71}  & 0.16 & 107K  & \textbf{90.28$\pm$3.66}    & 84.40$\pm$6.24    & 1.05 & 458K \\




\bottomrule
\end{tabular}
\caption{Dice score (\%) results of our \spac~ {($t$ indicates the $t$-th step)}  with other methods over all datasets. The running times of 3D registration is test on SLIVER dataset. Note that R2N2 works only for 2D registration. The official RCN and SYMNet are implemented only for 3D registration. }
\label{tab:registration}
\end{table*}

\subsubsection{Baselines.} 
This work focus on unsupervised deformable registration. We compare our method with several deep learning based image registration (DLIR) methods: VoxelMorph (VM)~\cite{balakrishnan2019voxelmorph}, VM-diff~\cite{dalca2019unsupervised}, SYMNet~\cite{mok2020fast}, R2N2~\cite{NEURIPS2019_dd03de08} and RCN~\cite{zhao2019recursive}. VM employs a U-Net structure with NCC loss to learn deformable registration, and VM-diff is a probabilistic diffeomorphic variant of VM. SYMNet is a state-of-the-art single-pass 3D registration method. R2N2 and RCN are sequence-based methods for 2D and 3D registration, respectively. To illustrate the effectiveness of the proposed method, we use the same network structure as VM. We also compare with two top-performing conventional registration algorithms, SyN~\cite{avants2008symmetric} and Elastix~\cite{klein2009elastix} with B-Spline~\cite{rueckert1999nonrigid}. In addition, we provide the result of using SSIM as the reward function.
We use Dice score as the evaluation metric.

\begin{figure}[t]
\centering
\setlength{\abovecaptionskip}{0.09in}
\setlength{\belowcaptionskip}{-0.05in}
\includegraphics[scale=0.45]{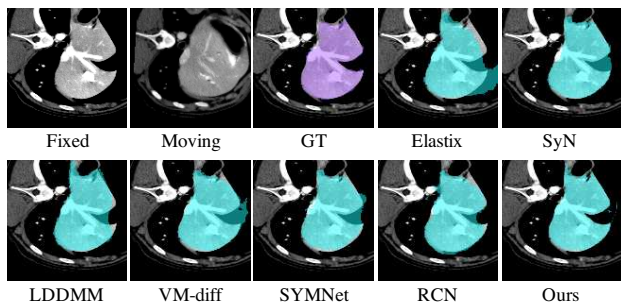}
\caption{Visual results of our \spac~with other methods on 3D liver dataset.  The warped moving image obtained by our \spac~is closer to the ground truth.
}
\label{fig:reg-liver}
\end{figure}

\subsection{Experimental Results}
\subsubsection{MNIST Digits Transform.} 
Figure \ref{mnist_box} shows the representative results and average Dice scores on MNIST digits transforms. In testing, the moving images are randomly scaled and rotated, which results in a larger and more challenging deformation field. Our method outperforms VM over all kinds of digits with significant improvements quantitatively and qualitatively, which also demonstrates that the proposed method has better generalizability and can work well on image pairs with large deformations.

\begin{figure}[t]
\setlength{\abovecaptionskip}{0in}
\setlength{\belowcaptionskip}{-0.1in}
\centering

\subfigure{
\includegraphics[width=0.23\textwidth]{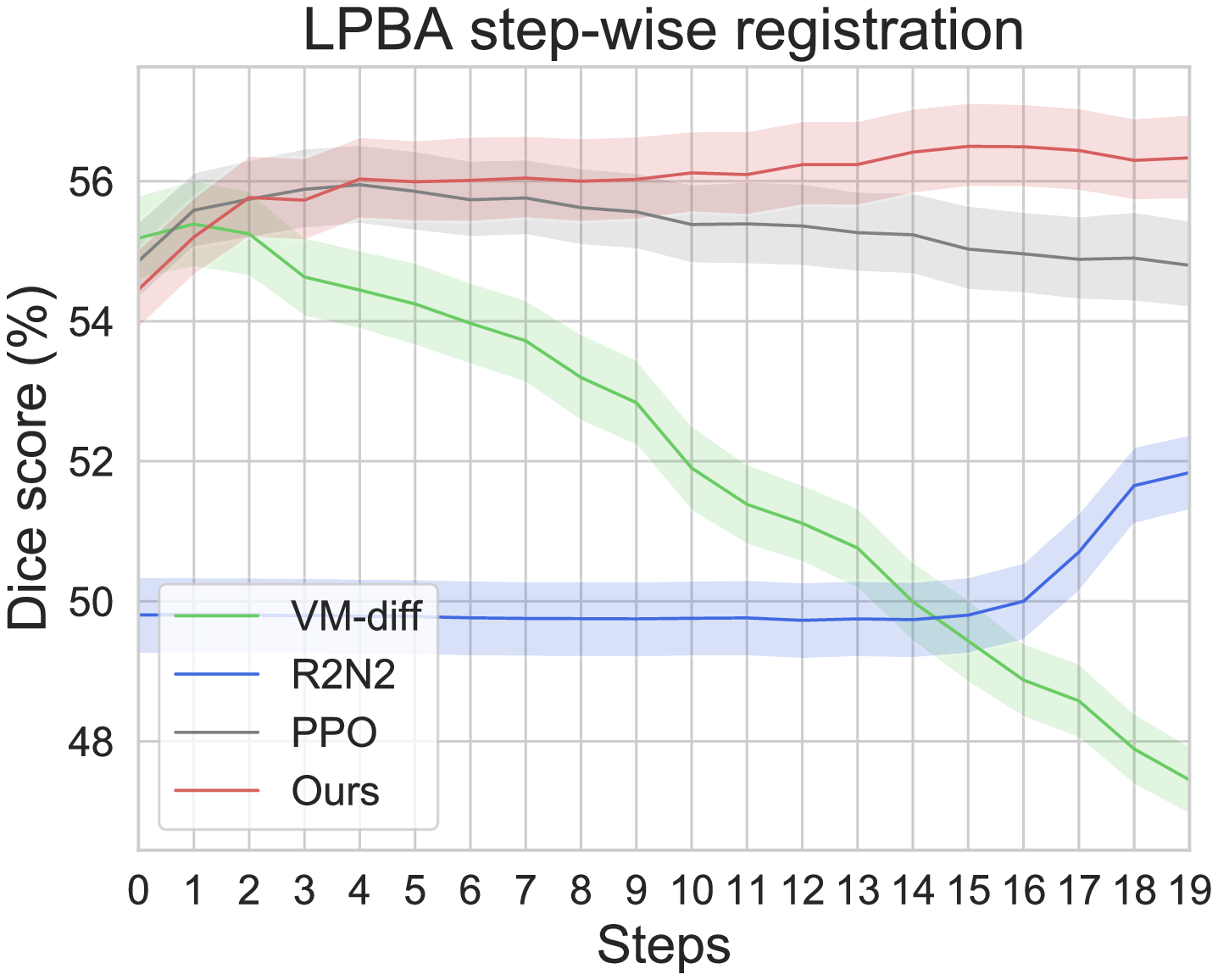}
\includegraphics[width=0.23\textwidth]{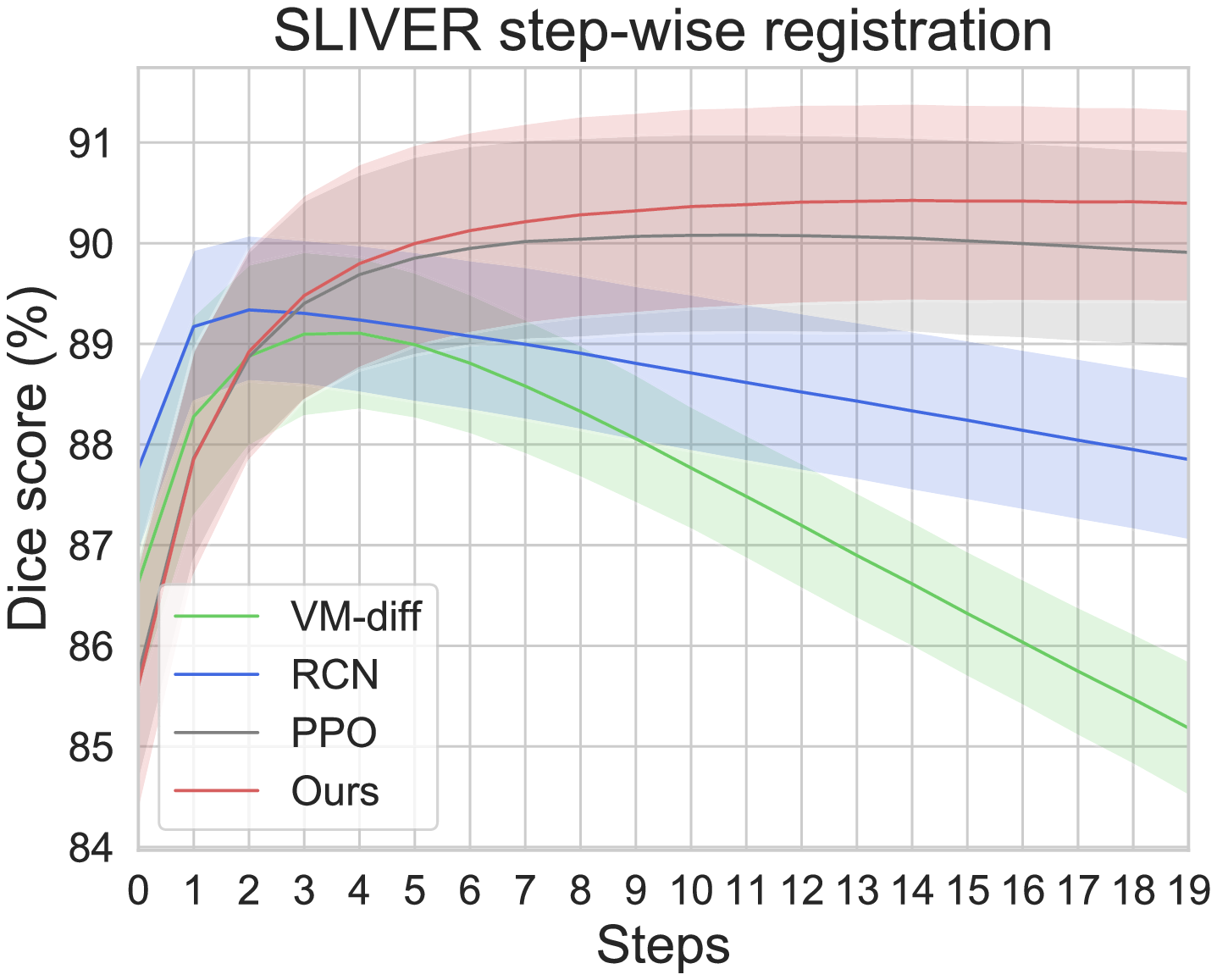}
}

\caption{Step-wise registration results overall datasets. The RL-based methods (Our \spac~ and PPO) perform more stable than other DL-based methods, and our \spac~ achieved the best performance. }
\label{fig:step_wise_curves}
\end{figure}

\begin{figure*}[t]
\centering
\includegraphics[scale=0.75]{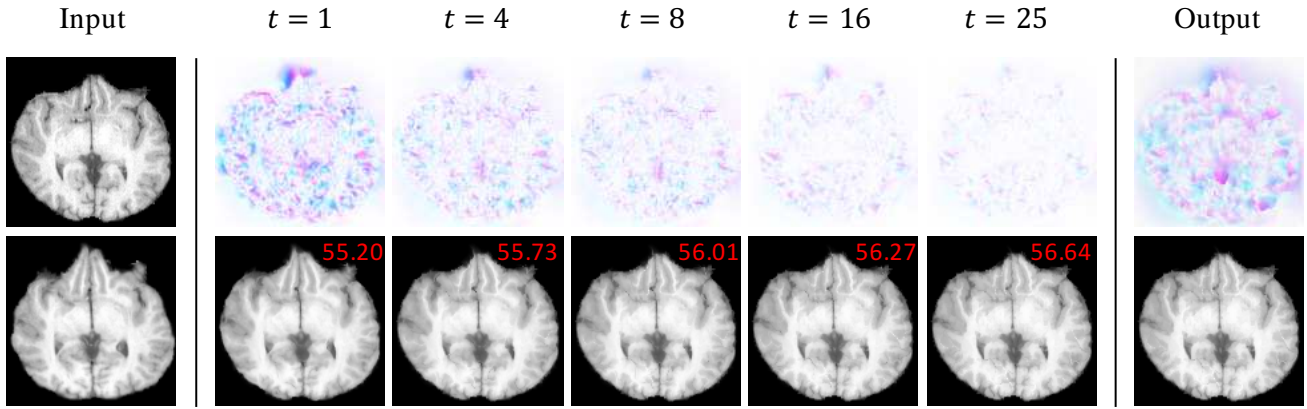}
\caption{A step-wise registration example of our method on the LPBA dataset. Top row is the visualized displacement field, where deep color represents a large deformation. The Dice score (red) keeps increasing along the steps. }
\label{fig:reg-multi}
\end{figure*}

\begin{figure}[ht]
\centering
\includegraphics[scale=0.5]{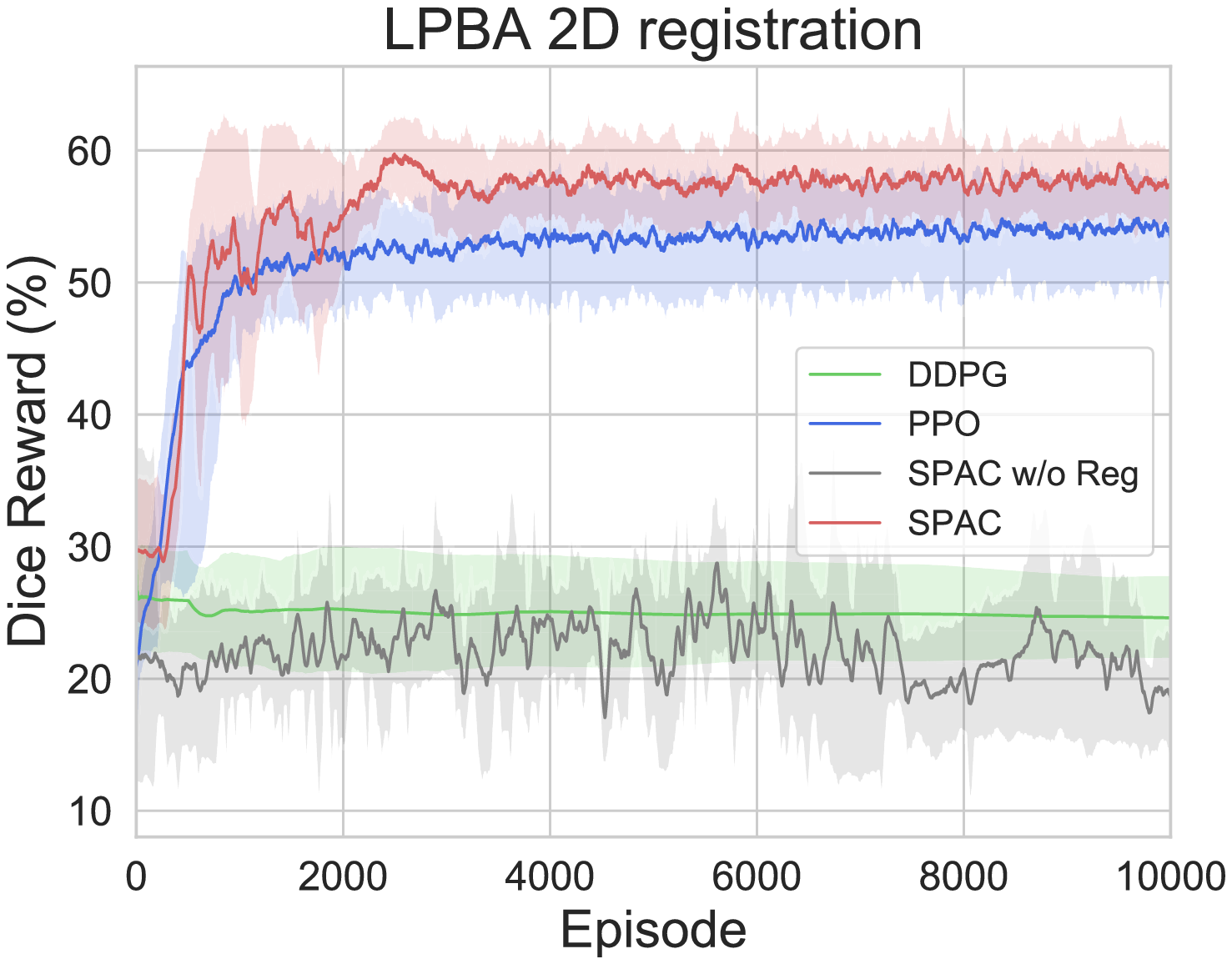}
\caption{Learning curves of several RL-based methods on LPBA dataset.}
\label{fig:rl_curve}

\end{figure}

\subsubsection{Medical Image Registration.}
Table \ref{tab:registration} summarizes the performance of our method with other state-of-the-art methods. The \spac~outperforms other methods over all 2D and 3D datasets.
Moreover, the LSPIG dataset has large deformation fields and is quite different from the training dataset (LiTS) in structure and appearance. The quantitative results on LSPIG illustrate that our framework works well in large deformation dataset and has better generalizability than other conventional DL-based methods. Note that our method performs registration in a step-wise manner, which results in a slower speed than most one-step methods such as VM and SYMNet. But \spac~is still faster than other multi-step methods such as R2N2 and RCN. The SSIM reward based \spac~ achieve good result on SLIVER but performs worse on LSPIG dataset compared with Dice reward based \spac~.
We visualized an example of registration results on the LSPIG dataset by overlaying the warped moving segmentation map on the fixed image in Figure \ref{fig:reg-liver}. The result shows that our model successfully learns registration even with a challenging, large deformation. The \spac~exceeds the performance of another step-wise method RCN, which demonstrates the effectiveness of our framework.

\subsection{Analysis}

\subsubsection{Step-wise registration.} The key idea of our method is to decompose the monolithic registration process into small steps by a lighter-weight CNN and progressively improves the transformed results. Figure \ref{fig:reg-multi} shows an example of step-wise registration process. The deformation fields visualized on the upper row illustrate that our method predicts transformation from coarse to the local refines step by step. We compared our method with PPO~\cite{schulman2017proximal} and other DL-based methods using step-wise registration in Figure \ref{fig:step_wise_curves}. As the registration step increases, the performance of DL-based methods gets worse, while the RL-based methods are more stable. The Dice score of \spac~is increasing all the time on LPBA and SLIVER datasets.

\begin{table}[t]
\centering
\scalebox{0.92}{
\begin{tabular}{l|ccc}
\toprule

& LPBA     &  SLIVER   & LSPIG  \\ \midrule

PPO-modified    & 55.82$\pm$3.49    & 89.30$\pm$3.63       & 83.55$\pm$6.24  \\

\spac-action    & 55.58$\pm$3.70    & 88.75$\pm$3.69       & 81.80$\pm$7.51  \\

\spac~w/o RL   & 54.89$\pm$3.80   & 85.43$\pm$4.14    & 80.72$\pm$7.34  \\

\spac~w/o Reg    & 44.67$\pm$3.74   & 79.34$\pm$4.02      & 72.45$\pm$6.25  \\

\spac~    &  \textbf{56.57$\pm$3.71  }   & \textbf{90.28$\pm$3.66 }      & \textbf{84.40$\pm$6.24 } \\

\bottomrule
\end{tabular}}
\captionof{table}{Dice score (\%) over several variants of our methods. `\spac-action' indicates that the critic evaluates the actor's action instead of planner in \spac. 
}
\label{tab:rl-dl}
\end{table}

\subsubsection{Compare with other RL methods.} To demonstrate our method in the reinforcement learning side. We modify our framework with other popular RL algorithms such as PPO \cite{schulman2017proximal} and DDPG \cite{lillicrap2015continuous} in Planner-Critic learning process. Moreover, we compare with the method discarding DL-based unsupervised registration loss (\spac~w/o Reg). The qualitative result of PPO-modified is shown in Table \ref{tab:rl-dl}. The training curves of each method are shown in Figure \ref{fig:rl_curve}. Our \spac~achieves batter performance than PPO. And the DDPG which uses deterministic policy is failed to convergence. The results also indicate that the RL agent can hardly deal with the DIR problem without unsupervised registration loss.

\subsubsection{Ablation Experiments.} We study the effect of some important settings in our framework, such as reinforcement learning, unsupervised registration learning, and evaluating plan with the critic. Note that in the settings without using registration loss, we evaluate the deformation field as the only action, and both the planner and actor are trained with the RL objective. As summarized in Table \ref{tab:rl-dl}, the result is unsatisfactory if we train \spac~without reinforcement learning, and it becomes worse if the training discards unsupervised registration loss. Critic evaluates actor's action (\spac-action) results in an inferior performance compared with the \spac~(critic evaluates planner). 

\section{Conclusion}
In this paper, a step-wise registration network based on reinforcement learning is proposed to handle large deformation problem which is especially observed in soft tissues. This method, \spac, an off-policy actor-critic model, can efficiently learn good policies in spaces with high-dimensional continuous actions and states. 
Central to \spac~is the proposed component `plan' which is defined in latent subspace and can guide the actor to generate high-dimensional actions. 
To the best of our knowledge, we are the first to propose a pure RL model to deformable medical image registration. Experiments based on diverse medical image datasets demonstrate that this architecture achieves significant gains over state-of-the-art methods, especially for the case with large deformations. With the superiority of good performance, we expect that the proposed architecture can potentially be extended to all deformable image registration tasks. 

\section*{Acknowledgements}
This work was supported in part by the National Natural Science Foundation of China under Grant 61602065, Sichuan province Key Technology Research and Development project under Grant 2021YFG0038.

\bibliography{aaai22}

\end{document}